\documentstyle[emulateapj,danonecolfloat]{article}

\def\beq#1 {\begin{equation}\label{#1}}
\def\eeq   {\end{equation}}
\def\beqa#1{\begin{eqnarray}\label{#1}}
\def\eeqa  {\end{eqnarray}}
\def\eq#1  {equation~(\ref{#1})}
\def\Eq#1  {Equation~(\ref{#1})}
\def\eqn#1 {~(\ref{#1})}

\def\sec#1{\S\ref{#1}}

\def\xbf {{\bf x}}

\def\Xbf {{\bf X}}
\def\Xbft{{\bf X}^T}

\def\ybf {{\bf y}}

\def\nbf {{\bf n}}

\def\Cbf {{\bf C}}

\def\rbf   {{\bf r}}

\def\rbfHat{\widehat{\rbf}}

\def\dT  {\Delta T}

\def\albf    {{\bf a}}
\def\albfHat {\widehat{{\bf a}}} 
\def\da      {\Delta\albfHat}

\def\bbf {{\bf b}}

\def\syn{synchrotron}

\def\ha    {$H_{\alpha}$}
\def\microk{$\mu$K}
\def\microm{$\mu$m}
\def\um{$\mu{\rm m}$}
\def\mj{MJy/sr}
\def\l {\ell}

\def\sn  {$S/N$}
\def\etal{{\frenchspacing\it et al.~}}

\def\ie  {{\frenchspacing\it i.e.}}

\def\ith{i^{th}}
\def\jth{j^{th}}
\def\expec#1{\langle#1\rangle}
\def\SS{{\bf\Sigma}}

\def\spose#1{\hbox to 0pt{#1\hss}}
\def\simlt{\mathrel{\spose{\lower 3pt\hbox{$\mathchar"218$}}
     \raise 2.0pt\hbox{$\mathchar"13C$}}}
\def\simgt{\mathrel{\spose{\lower 3pt\hbox{$\mathchar"218$}}
     \raise 2.0pt\hbox{$\mathchar"13E$}}}
\def\simpropto{\mathrel{\spose{\lower 3pt\hbox{$\mathchar"218$}}
     \raise 2.0pt\hbox{$\propto$}}}

\def\rn{\noindent\parshape 2 0truecm 8.8truecm 0.3truecm 8.5truecm}
\def\nn#1 #2{#1, #2.}				% Name with 1 initial
\def\nnn#1 #2 #3{#1, #2. #3.}			% Name with 2 initials
\def\nnnn#1 #2 #3 #4{#1, #2. #3. #4.}		% Name with 3 initials
\def\nnnnn#1 #2 #3 #4 #5{#1, #2. #3. #4. #5.}	% Name with 4 initials
\def\rg#1;#2;#3;#4;#5;#6 {\par\rn#1 #2, {\it #3}, {\bf #4}, #5 (``#6'') \par}
\def\rf#1;#2;#3;#4;#5 {\par\rn#1 #2, {\it #3}, {\bf #4}, #5\par}
\def\rfbook#1;#2;#3;#4;#5 {{\frenchspacing\par\rn#1 #2, {\it #3} (#4: #5)\par}}
\def\rfproc#1;#2;#3;#4;#5;#6 {{\frenchspacing\par\rn#1 #2, in {\it #3}, ed. #4 (#5: #6)\par}}
\def\rfprep#1;#2;#3  {{\par\rn#1 #2, #3\par}}
\def\rfprepp#1;#2;#3 {{\par\rn#1 #2, #3\par}}

\def\bfig{\begin{figure}[h] \centerline{\hbox{}}\vfill}
\def\efig{\end{figure}\vfill\newpage}

\def\fig#1{Figure~\ref{#1}}
\def\Fig#1{Figure~\ref{#1}}
 \journalid{337}{15 January 1989}
 \articleid{11}{14}

\slugcomment{March 6, 2000; Submitted to ApJL}
\begin{document}
\twocolumn[%%% Begin front material

\title{Galactic Contamination in the QMAP Experiment}

\author{Angelica de Oliveira-Costa
\footnote{
	 University of Pennsylvania, Dept. of Physics \& Astronomy, 
	 Philadelphia, PA 19104; 
%	 angelica@physics.upenn.edu
	 }$^{,}$
\footnote{Institute for Advanced Study, Olden Lane, Princeton, NJ 08540 }$^{,}$
\footnote{Princeton University, Department of Physics, Princeton, NJ 08544},
  Max Tegmark$^{a,b}$,
 Mark J. Devlin$^{a}$,
L.M. Haffner\footnote{Astronomy Department, University of Wisconsin, Madison, WI 53706}
     Tom Herbig$^{c}$,
Amber D. Miller$^{c}$, 
  Lyman A. Page$^{c}$,
Ron J. Reynolds$^{d}$,
    S.L. Tufte\footnote{Physics Department, Lewis and Clark College, Portland, OR 97219} 
}

% \author{Angelica de Oliveira-Costa\altaffilmark{1,2,3},
%         Max Tegmark               \altaffilmark{1,2}, 
%         Mark J. Devlin            \altaffilmark{1},
% 	Haffner, L.               \altaffilmark{4}
%         Tom Herbig                \altaffilmark{3}, 
%         Amber D. Miller           \altaffilmark{3},
%         Lyman A. Page             \altaffilmark{3},
% 	Ron Reynolds              \altaffilmark{4},
% 	S. L. Tufte               \altaffilmark{4}}
% 
% \altaffiltext{1}{University of Pennsylvania, Department of Physics \& Astronomy, Philadelphia, PA 19104;
%                  angelica@physics.upenn.edu} 
% \altaffiltext{2}{Institute for Advanced Study, Olden Lane, Princeton, NJ 08540} 
% \altaffiltext{3}{Princeton University, Department of Physics, Princeton, NJ 08544} 
% \altaffiltext{4}{} 
% 
%%%%%%%%%%%%%%%%%%% ABSTRACT: %%%%%%%%%%%%%%%%%%%%%%%%%%%

\begin{abstract}
We quantify the level of foreground contamination in the 
QMAP Cosmic Microwave Background (CMB) data 
with two objectives:
(a) measuring the level to which the QMAP power spectrum measurements 
need to be corrected for foregrounds and
(b) using this data set to further refine current foreground models.
We cross-correlate the QMAP data with a variety of foreground
templates.
The 30 GHz Ka-band data is found to be significantly correlated 
with the Haslam 408~MHz and Reich and Reich  
1420~MHz synchrotron maps, but not with the Diffuse 
Infrared Background Experiment (DIRBE) 240, 140 and 100{\microm} maps
or the Wisconsin H-Alpha Mapper (WHAM) survey.
The 40 GHz Q-band has no significant template correlations.
We discuss the constraints that this places on synchrotron, 
free-free and dust emission.
We also reanalyze the foreground-cleaned Ka-band data
and find that the two band power measurements 
are lowered by 2.3\% and 1.3\%, respectively.
% 1-29.7/30.4 = 2.3026%
% 1-37.5/38.0 = 1.3158%

\end{abstract}

]%%% End front material

%%%%%%%%%%%%%%%%%%% TEXT: %%%%%%%%%%%%%%%%%%%%%%%%%%%%%%%

\section{INTRODUCTION}

Quantifying Galactic emission in a cosmic microwave background (CMB) 
map is interesting for two different reasons. On one hand, 
the CMB is known to be a gold mine of information about 
cosmological parameters. Taking full advantage of this
requires accurate modeling and subtraction of Galactic 
foreground contamination.
On the other hand, the high fidelity maps being produced as
part of the current CMB gold rush offer a unique 
opportunity for secondary non-CMB science. This includes a greatly 
improved understanding of Galactic emission processes
between 10 and $10^3$ GHz.

This paper is motivated by both of these reasons.
The QMAP experiment (Devlin {\etal} 1998; 
Herbig {\etal} 1998; 
de Oliveira-Costa \etal 1998b, hereafter dOC98b)
is one of the CMB experiments that has produced
a sky map with accurately modeled noise properties,
lending itself to a cross-correlation analysis 
with a variety of foreground templates.
We present such an analysis in \sec{CorrSec} and \sec{basicresults},
then compute the corresponding correction to the
published QMAP power spectrum measurements in 
\sec{CMBsec} and finish by discussing the implications for 
Galactic foreground modeling in \sec{ForegSec}.

%\section{Data analysis and basic results} 
\section{METHOD}
\label{CorrSec}

The multi-component fitting method that we use was presented in
detail in de Oliveira-Costa \etal 1999 (hereafter dOC99), 
so we review it only briefly here.
The joint QMAP map from both flights 
consists of $N=3164$ (Ka-band, 26 to 36~GHz) and 
$4875$ (Q-band, 36-46~GHz)
measured sky temperatures (pixels) $y_i$.
We model this map as a sum of CMB fluctuations $x_i$, 
detector noise $n_i$ 
and $M$ Galactic components 
whose spatial distributions are traced in part by external 
foreground templates. Writing these contributions as 
$N$-dimensional vectors, we obtain
\beq{signals}
\ybf = \Xbf\albf + \xbf + \nbf,
\eeq
where $\Xbf$ is an $N\times M$ matrix whose rows
contain the various foreground templates convolved with the 
QMAP beam (\ie, $\Xbf_{ij}$ would be the $\ith$
observation if the sky had looked like the $\jth$ foreground 
template), and $\albf$ is a vector of size $M$ that gives the 
levels at which these foreground templates are present in the 
QMAP data.

We treat $\nbf$ and $\xbf$ as uncorrelated random
vectors with zero mean and the $\Xbf$ matrix as constant,
so the data covariance matrix is given by
\beq{varCMB}
  \Cbf \equiv 
       \expec{\ybf \ybf^T} - \expec{\ybf} \expec{\ybf^T} =
       \expec{\xbf \xbf^T} +  \expec{\nbf \nbf^T},
\eeq
where
\beq{corrFunc}
	\expec{\xbf \xbf^T}_{ij} \equiv 
	\sum_{\l=2}^\infty \frac{2\l+1}{4\pi}
	P_\l(\rbfHat_i \cdot \rbfHat_j) W_\l^2 C_\l 
\eeq
is the CMB covariance matrix and $\expec{\nbf \nbf^T}$ is 
the QMAP noise covariance matrix (a detailed description 
of the calculation of $\expec{\nbf \nbf^T}$ is presented 
in dOC98b).
We use a flat power spectrum $C_\l\propto 1/\l(\l+1)$
normalized to a $Q_{flat}=(5C_2/4\pi)^{1/2}=30$~{\microk}
(dOC98b). We model the QMAP beam as a Fisher function 
%(Fisher {\etal} 1987) 
with FWHM$=\sqrt{8\ln 2}\sigma=0.9^\circ$ 
for Ka-band and $0.6^\circ$ for Q-band, which gives a 
$W_\l\approx e^{-\sigma^2\l(\l+1)/2}$.

Since our goal is to measure $\albf$, both 
$\xbf$ 
and 
$\nbf$ 
act as unwanted noise in equation (\ref{signals}).
Minimizing
$ \chi^2 \equiv 
          (\ybf - \Xbf\albf)^T 
          {\bf C}^{-1}
          (\ybf - \Xbf\albf) $
yields the minimum-variance estimate of $\albf$,
\beq{alpha}
   \albfHat  = 
   \left[\Xbft ~ {\bf C}^{-1} ~  \Xbf\right]^{-1}
   \Xbft ~ {\bf C}^{-1} ~  \ybf
\eeq
with covariance matrix
\beq{varalpha}
   {\SS}\equiv\expec{\albfHat^2} - \expec{\albfHat}^2 =
   \left[\Xbft ~ {\bf C}^{-1} ~ \Xbf\right]^{-1}.
\eeq
The error bars on individual correlations are therefore 
$\Delta \widehat{a}_i=\SS_{ii}^{1/2}$. This includes the 
effect of chance alignments between the CMB and the  
various template maps, since the CMB anisotropy term is 
incorporated in $\expec{\xbf\xbf^T}$.

%\subsection{Correlations \& Variances}\label{corrANDvar}
\section{BASIC RESULTS}
\label{basicresults}

\begin{figure}[tb] 
\centerline{\epsfxsize=9cm\epsfbox{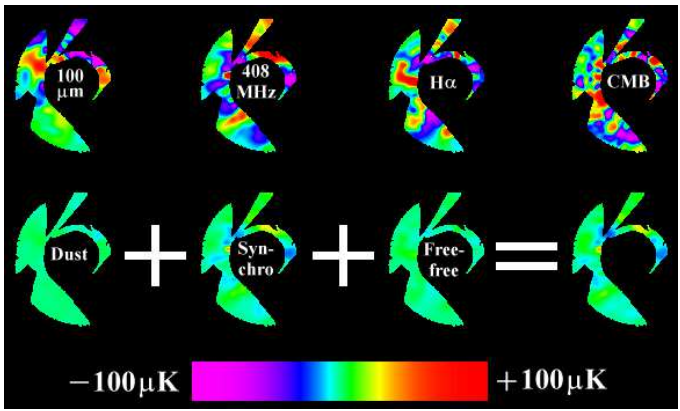}}
\caption{
\label{MapFig}\footnotesize%
Top: from left to right, the
100\microm, Has, \ha~ and QMAP Ka-band maps are shown,
each in their own units. They are centered roughly on
the North Celestial Pole.
Bottom: from left to right, the three templates above are
multiplied by the $\albfHat$-values from Table 1 and summed.
These four maps and the CMB map all use the
{\microk} scale shown by the color bar, so the
the best fit map of the foreground contribution 
(lower right) is seen to be quite subdominant 
to the CMB signal (upper right).
To make the CMB plot less noisy, 
the Wiener filter described in dOC98b has been applied to all 
eight maps before plotting.
}
\end{figure}

We cross-correlate the QMAP data with two different \syn~ 
templates: the 408~MHz survey (Haslam \etal 1982) and the 
1420~MHz survey (Reich 1982; Reich and Reich 1986), hereafter Has and 
R\&R, respectively. To study dust and/or free-free emission, 
we cross-correlate the QMAP data with three Diffuse Infrared 
Background Experiment (DIRBE) sky maps at wavelengths 100, 
140 and 240\um~ (Boggess \etal 1992) and with the Wisconsin 
H-Alpha Mapper (WHAM)\footnote{
		Details also at {\it http://www.astro.wisc.edu/wham/}.}
survey (Haffner \etal 1999).
For definiteness, we use the DIRBE 100{\um} channel when 
placing limits below since it is the least noisy of the 
three DIRBE channels. 
Three of our templates are shown together with the 
Ka-band QMAP map in \fig{MapFig}.
Most of our interesting results come
from the QMAP Ka-band, since the Q-band was substantially noisier
(the opposite was true for the Saskatoon experiment -- 
see de Oliveira-Costa \etal 1997, hereafter dOC97).

Before calculating the correlations, we convolve 
the template maps with the QMAP beam function. We also remove
the monopole and dipole from both the 
templates and the QMAP maps.
As a consequence, our results depend predominantly 
on the small scale intensity variations in the templates and 
are insensitive to the zero levels of the QMAP data and the 
templates. 

% \medskip
% \centerline{{\vbox{\epsfxsize=9cm\epsfbox{foregfig.eps}}}}
% \noindent{\small
% \label{maps}
% Fig.~1 --- 
% Top: from left to right, the
% 100\microm, Has, \ha~ and QMAP Ka-band maps are shown,
% each in their own units. They are centered roughly on
% the North Celestial Pole.
% Bottom: from left to right, the three templates above are
% multiplied by the $a$-values from table 1 and summed.
% These four maps and the CMB map all use the
% {\microk} scale shown by the color bar, so the
% the best fit map of the foreground contribution 
% (lower right) is seen to be quite subdominant 
% to the CMB signal (upper right).
% To make the CMB plot less noisy, 
% the Wiener filter described in dO98 is has been applied to all 
% eight maps before plotting.
% }
% \bigskip

Table~1 shows the coefficients $\albfHat$ and the corresponding 
fluctuations in antenna temperature in the QMAP data 
($\Delta T = \albfHat \sigma_{Gal}$, where $\sigma_{Gal}$ is 
the standard deviation of the template map). Statistically 
significant ($>2\sigma$) correlations are listed in boldface. 
Note that the fits are done jointly for $M=3$ templates. The DIRBE, 
Haslam and \ha~ correlations listed in Table~1 correspond to joint 
100\um$-$Has$-$\ha~ fits, whereas the R\&R numbers correspond to 
a joint 100\um$-$R\&R$-$\ha~ fit.
Only the two \syn~ templates are found to be correlated with the 
Ka-band, while no correlations are found for the Q-band. 
Repeating the
analysis done for two different Galactic cuts (20$^{\circ}$ and 
30$^{\circ}$) indicates
%shows 
that the bulk of this contamination is at
latitudes lower than 30$^{\circ}$. 

\begin{table}
{\label{CorrTab}\footnotesize%
\caption{\centerline{\footnotesize Correlations for Ka- and Q-band data.}}
%\vspace{-0.1cm}
\begin{center}
\begin{tabular}{llrrrrc}
\hline  
& & & & & & \\
\multicolumn{1}{c}{$b$ \& $\nu$}            &
\multicolumn{1}{c}{Maps$^{(a)}$}            &
\multicolumn{1}{r}{$\albfHat\pm\da^{(b)} $} & 
\multicolumn{1}{r}{${\albfHat\over\da}   $} &    
\multicolumn{1}{c}{$\sigma_{Gal}^{(c)}   $} &          
\multicolumn{1}{c}{$\dT$ [\microk]$^{(d)}$} &
\multicolumn{1}{c}{$\beta$}                 \\
& & & & & & \\
\hline 
%%%%%%%%%%%%%%%%%%%%%%%%%%%%%%%%%%%%%%%%%%%%%%%%%%%%%%%%%%%%%%%%%%%%%%%%%%%
\hline 
All Sky		   &100\microm  &-3.5 $\pm$ 6.9  &     -0.5   &  3.1  &-10.8$\pm$21.4  &      \\
Ka-band		   &Has         & 6.6 $\pm$ 2.6  &{\bf  2.5}  &  4.3  & 28.4$\pm$11.2  &-2.8  \\
		   &R\&R        & 0.3 $\pm$ 0.1  &{\bf  2.6}  &141.4  & 42.4$\pm$14.1  &-2.6  \\
		   &\ha         &14.6 $\pm$ 9.5  &      1.5   &  1.0  & 14.6$\pm$ 9.5  &      \\
		   & & & & & & \\
%%%%%%%%%%%%%%%%%%%%%%%%%%%%%%%%%%%%%%%%%%%%%%%%%%%%%%%%%%%%%%%%%%%%%%%%%%%%
All Sky		   &100\microm  &-1.7 $\pm$ 5.4  &     -0.3   &  3.9  & -6.7$\pm$21.4  &      \\
Q-band		   &Has         & 0.7 $\pm$ 2.8  &      0.2   &  6.1  &  4.2$\pm$17.2  &-2.7  \\
		   &R\&R        &-0.05$\pm$ 0.09 &     -0.6   &182.3  & -9.4$\pm$16.3  &-3.0  \\
		   &\ha         &-6.0 $\pm$11.2  &     -0.5   &  1.1  & -6.3$\pm$11.8  &      \\
		   & & & & & & \\
%%%%%%%%%%%%%%%%%%%%%%%%%%%%%%%%%%%%%%%%%%%%%%%%%%%%%%%%%%%%%%%%%%%%%%%%%%%%
$b>20^{\circ}$     &Has         & 6.1 $\pm$ 2.7  &{\bf  2.3}  & 2.2   & 13.9$\pm$ 6.2  &-2.8  \\
Ka-band		   &R\&R        & 0.3 $\pm$ 0.1  &{\bf  2.2}  &82.5   & 24.0$\pm$10.9  &-2.7  \\
		   & & & & & & \\
%%%%%%%%%%%%%%%%%%%%%%%%%%%%%%%%%%%%%%%%%%%%%%%%%%%%%%%%%%%%%%%%%%%%%%%%%%%%
$b>30^{\circ}$     &Has         & 4.1 $\pm$ 5.2  &      0.8   & 1.4   &  5.8$\pm$ 7.4  &-2.9  \\
Ka-band		   &R\&R        & 0.4 $\pm$ 0.3  &      1.2   &28.6   & 10.5$\pm$ 8.9  &-2.6  \\
%%%%%%%%%%%%%%%%%%%%%%%%%%%%%%%%%%%%%%%%%%%%%%%%%%%%%%%%%%%%%%%%%%%%%%%%%%%%
\hline    
\end{tabular}
\end{center}     
}
%%%%%%%%%%%%%%%%%%
\vspace{+0.2cm}
\noindent{\small  
	$^{(a)}$ The DIRBE and Haslam correlations listed in this table 
	         correspond to joint 100\um$-$Has$-$\ha~ fits, whereas 
		 the R\&R numbers correspond to a joint 100\um$-$R\&R$-$\ha~ 
		 fit; \\
	$^{(b)}$ $\albfHat$ has units \microk (\mj)$^{-1}$ for the 100\microm~ 
	         template, \microk/K for the Has template, \microk/mK for 
		 the R\&R template and \microk/R for the \ha~template; \\ 
	$^{(c)}$ $\sigma_{Gal}$ has units of the templates;\\
	$^{(d)}$ $\dT \equiv (\albfHat\pm\da) \sigma_{Gal}$.
	}
\end{table}

% \subsection{Testing the Cross-Correlation Technique}

As in dOC97, de Oliveira-Costa 
\etal 1998a (hereafter dOC98a) and dOC99, the 
cross-correlation software was tested by analyzing constrained realizations 
of CMB and QMAP instrument noise. From 1000 realizations, 
we recovered unbiased estimates $\albfHat$ with a variance 
in excellent agreement with equation (\ref{varalpha}).  
As an additional test, we computed 
$\chi^2 \equiv (\ybf - \Xbf\albf)^T {\bf C}^{-1} (\ybf - \Xbf\albf)$
and obtained $\chi^2/N \approx 1$ in all cases.
Including a synchrotron template lowered $\chi^2$ by a significant amount
(18 for R\&R and 9 for Has), whereas adding the other templates 
resulted in insignificant reductions $\Delta\chi^2\sim 1$.

\section{IMPLICATIONS FOR CMB}
\label{CMBsec}

The lower right map in \fig{MapFig} shows 
$\Xbf\albf^T$, \ie, 
our best fit estimate
of the foreground contribution to the QMAP Ka-band.
To quantify the foreground contribution to the published
QMAP power spectrum measurements, we repeat the
exact same analysis described in dOC98b after subtracting out this map,
\ie, with the map $\ybf$ replaced 
by $\ybf - \Xbf\albf^T$.

\begin{figure}[tb] 
\vspace{-0.3cm}
\centerline{{\vbox{\epsfxsize=9.5cm\epsfbox{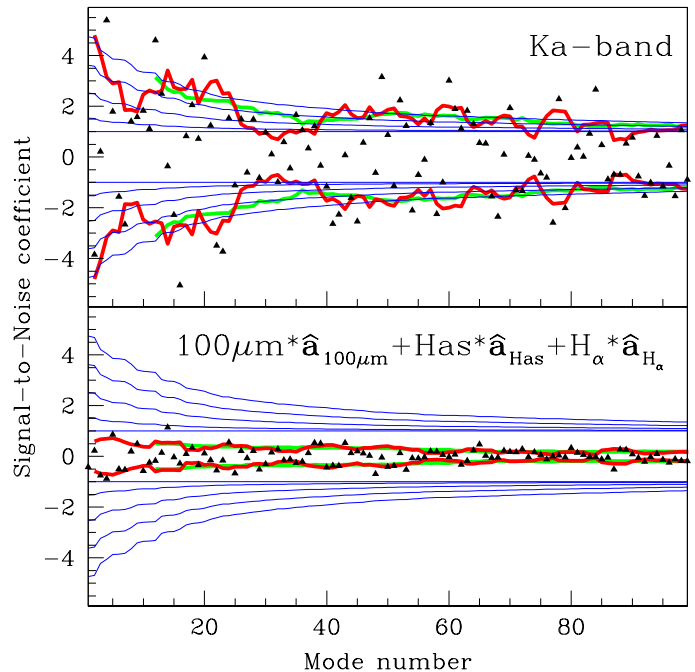}}}}
\vspace{-0.2cm}
\caption{
\label{SNfig}\footnotesize%
S/N eigenmode coefficients of the Ka-band data (top) and
combined foregrounds (bottom).
The wiggly curves show the a running average of the
rms fluctuations with bin length 5 and 25
The smooth curves show the theoretically expected 
rms for flat power spectra with
$Q_{flat}=$0, 10, 20, 30 and 40 $\mu$K.
}
\end{figure}

\begin{figure}[tb] 
\vspace{-1.2cm}
\centerline{{\vbox{\epsfxsize=9.5cm\epsfbox{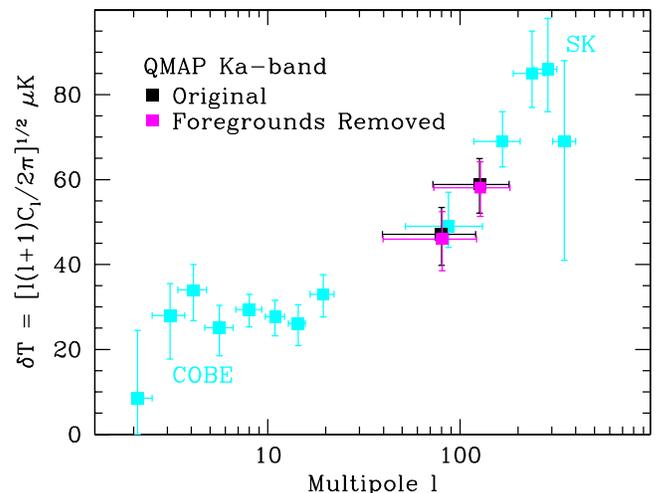}}}}
\vspace{-1.5cm}
\caption{
\label{ClFig}\footnotesize%
Power spectrum from QMAP experiment.
The foreground removed points (in magenta) is shown
with the original points (in black).
}
\end{figure}

The dOC98b band powers were computed
by expanding the map in signal-to-noise (\sn) eigenmodes 
(Bond 1995; Bunn and Sugiyama 1995), 
weight vectors $\bbf_i$ that solve the generalized 
eigenvalue equation
\beq{a17}
\expec{\xbf \xbf^T} \bbf_i = \lambda_i \expec{\nbf \nbf^T} \bbf_i.
\eeq
When the $\bbf$'s are sorted by decreasing eigenvalue $\lambda$, 
they tend to probe from larger to smaller angular 
scales. The $S/N$ expansion coefficients are shown in 
\fig{SNfig}, and are seen to be substantially smaller
for the foregrounds than for the CMB.
As described in dOC98b, we obtain a statistically independent power 
estimate from the square of each mode and 
then average these individual 
estimates with inverse-variance weighting to obtain 
the band power estimates listed in Table 2 and 
shown in \fig{ClFig}.
The Ka-band (30 GHz) band powers are seen to drop by less than a few 
percent when the foreground signal is subtracted,
whereas the Q-band (40 GHz) contamination is too small to quantify.
The Ka-band correction is slightly smaller on small angular scales:
1.3\% instead of 2.3\%.
This is expected, since diffuse Galactic foregrounds
are expected to have a redder power spectrum 
than the CMB.

% \subsection{Pointing}

As a side benefit, our statistically significant detection of foregrounds
allowed an independent confirmation of the QMAP pointing solution.
We reanalyzed the QMAP data set with the pointing solution 
offset by $\pm 0.5^{\circ}$ in azimuth, and found the highest 
correlations for the original pointing.

\begin{table}
\caption{\label{PowerTab}\footnotesize%
The angular power spectrum
before and after foreground removal.
   The band powers $\delta T_\l\equiv[\l(\l+1)C_\l/2\pi]^{1/2}$ 
   have uncorrelated errors and window functions whose mean and 
   rms width are given by $\l_{eff}$ and $\Delta\l$
   just as in dOC98b.}
%   \vspace{-0.5cm}
   \begin{center}
   {\footnotesize
   \begin{tabular}{lcrccc}
   \hline
   \hline
   \multicolumn{1}{l}{Flight}        &
   \multicolumn{1}{l}{Band}          &
   \multicolumn{1}{c}{$\l_{eff}$}    & 
   \multicolumn{1}{c}{$\Delta \l $}  & 
   \multicolumn{1}{c}{$\delta T$ before}    &
   \multicolumn{1}{c}{$\delta T$ after}    \\
   \hline
%%%% OLD DATA:
%   1+2  &Ka	&80	&41  &$47^{+6}_{-7}$  &$47^{+6}_{-7}$\\
% 	 &Ka	&126	&54  &$59^{+6}_{-7}$  &$59^{+6}_{-7}$\\
%%%%%%%%%%%%%%
    1+2  &Ka	&81	&41  &47$^{+6}_{-7}$  &46$^{+6}_{-8}$\\
    1+2	 &Ka	&127	&55  &59$^{+6}_{-7}$  &58$^{+6}_{-7}$\\
\hline
   \end{tabular}
   }
   \end{center}
\end{table}

\begin{figure}[tb] 
%\vspace{-1.2cm}
\centerline{{\vbox{\epsfxsize=9.5cm\epsfbox{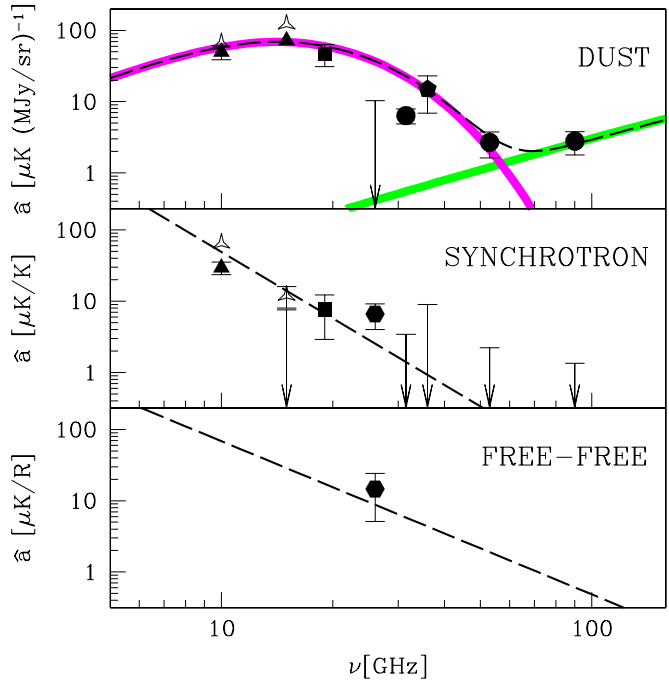}}}}
%\vspace{-1.5cm}
\caption{
\label{SpectrumFig}\footnotesize%
Galactic contaminants as function of frequency.
Frequency dependence of DIRBE-correlated emission (top),
Haslam-correlated emission (middle), and
\ha-correlated emission (bottom).
From left to right, the dust points are
from Tenerife 10 GHz and 15 GHz (dOC99: filled triangles;
Mukherjee et. al. 2000 Table 2 $b>20^{\circ}$: open triangles), 
19 GHz (dOC98a), this work (the upper limit),
DMR 31 GHz (Kogut et al. 1996), Saskatoon (dOC97), 
DMR 53 and 90 GHz (Kogut et. al. 1996).
The Synchrotron points are from the
same experiments as the dust points at the 
corresponding frequencies.
The entry in the bottom plot are
from this work.
Upper limits are 2-$\sigma$. 
The dashed model curves show the frequency dependence
characterizing (spinning + vibrating) dust
(top), synchrotron radiation (middle) and free-free emission 
(bottom). The normalization of the synchrotron model
$a\propto\nu^{-\beta}$ is fixed by the fact that, by definition,
$a=1$K$/\mu$K$=10^6$ at 408 MHz, and the dust model
must similarly reproduce DIRBE at 100$\mu$m.
}
\end{figure}

\section{IMPLICATIONS FOR FOREGROUND MODELING}
\label{ForegSec}

\Fig{SpectrumFig} shows a compilation of measured correlations between 
foreground templates and CMB maps at various frequencies.
Comparisons are complicated by the fact that 
the foreground level
$\Delta T = \albfHat \sigma_{Gal}$ depends strongly on 
galactic latitude via $\sigma_{Gal}$. We therefore 
plot $\albfHat$ instead, \ie, the factor giving the
frequency dependence of emission per unit foreground.
Such measurements were used to normalize recent foreground
models such as those of Bouchet \& Gispert (1999)
and Tegmark {\etal} (1999).
Below we discuss how our QMAP results affect such models.

\subsection{Synchrotron}

Writing the frequency dependence as $a\propto\nu^{\beta}$
and recalling that the correlation coefficients are, by definition, 
$a=1$K$/\mu$K$=10^6$ for Has at 408 MHz and 
$a=1$mK$/\mu$K$=10^3$ for R\&R at 1420 MHz,
we obtain the spectral index limits
$-2.7 \simlt \beta \simlt -3.3$ for the Ka--Has correlation
and $-2.6 \simlt \beta \simlt -2.8$ for the Ka--R\&R correlation.
These values are slightly steeper than the 
canonical sub-GHz slope of $-2.7\simlt \beta \simlt -2.9$
(Davies {\etal} 1998; Platania {\etal} 1998), but
consistent with a steepening of the spectrum of 
cosmic ray electrons at higher energies
(Rybicki and Lightman 1979).
%%%%%%%%%%%%%%%%%%%%%%%%%%%%%%%%%%%%%%%%%%%%%%%%%%%%%%%%%%%%%%%%%%%%%
% This result persisted when we repeated the analysis for three
% separate bins of Galactic latitude $b$ --- see \fig{LatitudeFig}.
%%%%%%%%%%%%%%%%%%%%%%%%%%%%%%%%%%%%%%%%%%%%%%%%%%%%%%%%%%%%%%%%%%%%%

The relatively high QMAP synchrotron signal seen in \fig{SpectrumFig} could
be interpreted as slight spatial variability of the
frequency dependence (Tegmark 1999), but may also have other explanations.
For instance, the worst striping problems in the Haslam map are right
around the North Celestial Pole, which may have caused Saskatoon
to underestimate the true synchrotron level there (dOC97).

\subsection{Spinning dust \& free-free emission}

An important question is whether the DIRBE-correlated
signal seen by so many experiments (see \fig{SpectrumFig}, top) is due to
dust-correlated free-free emission (Kogut \etal 1996) or 
spinning dust grains (Draine \& Lazarian 1998).
The turndown at low frequencies suggests a spinning dust
interpretation (dOC99), but an analysis using improved 
Tenerife data (Mukherjee \etal 2000) may have re-opened this 
question\footnote{
As seen in \fig{SpectrumFig}, the Mukherjee \etal results 
for $b>20^{\circ}$ also shown a turndown between 15 and 10~GHz.
Figure 7 in their paper looks different because it is an average
including data down to $b=0$.
This may still be consistent with spinning dust
being the worst foreground for the (Galaxy cut) MAP data,
since free-free emission is believed to be more concentrated
in the Galactic plane than dust.
}.
Our present results cannot settle the issue, but offer a few
additional clues.

If free-free emission is responsible for the correlation, then substantial 
\ha~ emission would be expected as well.
\Fig{SpectrumFig} (bottom) shows the expected correlation $a$
for the case of $8,000$K gas (Bennett \etal 1992).
It is seen that the \ha-correlation, although
marginal at best, is consistent with the theoretical curve. 
% Table 1 shows that our detected $H\alpha$-correlation
% is marginal at best, not significant at $2\sigma$ level.
Moreover, Table 1 shows that the limit on the dust-correlated
emission is low enough to be compatible with a free-free origin
(\ie, with the $\sim 15\mu$K $H\alpha$-correlated component). 

To clarify this issue, we computed the correlation 
between the dust and \ha~ maps.
As described in dOC99, 
eq. (\ref{varalpha}) shows that we can interpret 
$\SS^{-1}$ as the covariance between the various templates 
with dimensionless correlation coefficients
  $r_{ij} \equiv \SS^{-1}_{ij} (\SS^{-1}_{ii}\SS^{-1}_{jj})^{-0.5}$.
%% 
%Here we use the Ka-band pixel weights.
Like in dOC99, the DIRBE maps were found to be almost perfectly correlated,
and essentially uncorrelated ($r^2\simlt 3\%$) with the radio maps.
The Has and R\&R maps are correlated with
$r\sim 83\%$ for $b > 20^{\circ}$.
%% 
%% What is new here is the correlation between 
%% the DIRBE 100$\mu$m map and the \ha~ template. 
%% 
As a new result, we obtain 	
a marginal correlation $r\approx 0.2$
between the DIRBE maps and the \ha~ template. 
Since the statistical properties of these maps are not accurately
known, we computed error bars by repeating the analysis with
one of the templates replaced by $2\times 2\times 72=288$ 
transformed maps, rotated around the Galactic axis by 
multiples of 5$^{\circ}$ and/or flipped vertically 
and/or horizontally. 
The actual correlation was found to be larger than 85\% of
these, showing that
the correlation is not 
significant at the $2\sigma$ level:
$\albfHat = (0.25\pm 0.19)\,$R/MJy$\,$sr$^{-1}$ 
$(1\sigma)$.
%\footnote{ 
	This result is significantly smaller than that recently 
	found by Lagache \etal (2000) for their DIRBE--WHAM 
	correlation done in a different region of the sky,
	but compatible with other marginal dust--\ha~ correlations
	(McCullough 1997; Kogut 1997).
%}.  

This poor correlation is a challenge for the pure
free-free hypothesis, which maintains that 
microwave emission traces dust because dust traces free-free emission.
% (with only $r^2\sim 20\%$) 
A cross-correlation analysis with large 
frequency and sky coverage will hopefully be able
to unambiguously determine the relative levels of free-free and dust emission
in the near future.
 \medskip
We would like to thank Matias Zaldarriaga 
for encouraging us to complete it. 
Support for this work was provided by the Packard Foundation,
NSF grants PHY 92-22952 \& PHY 96-00015, and NASA grant
NAG5-6034. 
We acknowledge the NASA office of Space Sciences, the COBE flight 
team, and all those who helped process and analyze the DIRBE data.

%%%%%%%%%%%%%%%%%%%%%% REFERENCES: %%%%%%%%%%%%%%%%%%%%%%%%%

\end{document}